\documentclass[aps,prl,a4paper,twocolumn]{revtex4}
\pdfoutput=1

\usepackage{graphicx}
\usepackage{times}
\usepackage{amsmath}
\usepackage{amssymb}
\usepackage{textcomp}
\usepackage{bbold}
\usepackage{microtype}
\usepackage{etoolbox}

\usepackage{anysize}
\marginsize{1.4cm}{1.3cm}{0.3cm}{0.8cm}

\usepackage{color}
\definecolor{MyDarkBlue}{rgb}{0,0,0.75}
\usepackage[pdftex,pdfborder={0 0 0},citecolor=MyDarkBlue, urlcolor=MyDarkBlue,
  linkcolor=MyDarkBlue, colorlinks=true, bookmarksopen=true]{hyperref}

\hypersetup{pdfauthor={Helmut Strobel},pdftitle={Fisher Information and entanglement of non-Gaussian spin states}}

\newcounter{myequationcnt}
\makeatletter
\@addtoreset{equation}{myequationcnt}
\makeatother

\newcounter{myfigurecnt}
\makeatletter
\@addtoreset{figure}{myfigurecnt}
\makeatother

\newcommand{\ud}{\mathrm{d}}

\newcommand{\dHs}{d^2_{\rm H}}
\newcommand{\Frq}{\mathcal{F}}
\renewcommand{\i}{\mathrm{i}}

\date{December 24, 2013}

\begin{document}
\title{\Large{Fisher Information and entanglement of non-Gaussian spin states}}
\author{Helmut Strobel$^{1\ast}$, Wolfgang Muessel$^{1}$, Daniel Linnemann$^{1}$, Tilman Zibold$^{1}$, David B. Hume$^{1}$, Luca Pezz\`e$^{2}$, Augusto Smerzi$^{2}$ and Markus K. Oberthaler$^{1}$}
\affiliation{%
$^{1}$Kirchhoff-Institut f\"ur Physik, Universit\"at Heidelberg, Im Neuenheimer Feld 227, 69120 Heidelberg, Germany.
\\
$^{2}$QSTAR, INO-CNR and LENS, Largo Enrico Fermi 2, 50125 Firenze, Italy.
\\
$^\ast$Corresponding author. E-mail:  \href{mailto:fisherinformation@matterwave.de}{\rm FisherInformation@matterwave.de}.
}
\maketitle
%
%
%
%
%
\textbf{Entanglement is the key quantum resource for improving measurement sensitivity beyond classical limits. However, the production of entanglement in mesoscopic atomic systems has been limited to squeezed states, described by Gaussian statistics. Here we report on the creation and characterization of non-Gaussian many-body entangled states. We develop a general method to extract the Fisher information, which reveals that the quantum dynamics of a classically unstable system creates quantum states that are not spin squeezed but nevertheless entangled. The extracted Fisher information quantifies metrologically useful entanglement which we confirm by Bayesian phase estimation with sub shot-noise sensitivity. These methods are scalable to large particle numbers and applicable directly to other quantum systems.}

Multiparticle entangled states are the key ingredients for advanced quantum 
technologies~\cite{NielsenBOOK}. Various types have been achieved in experimental settings ranging from ion 
traps~\cite{BlattNATURE2008}, photonic 
systems~\cite{PanRMP2012} and solid state 
circuits~\cite{VlastakisSCIENCE2013} to Bose-Einstein condensates. For the latter, squeezed 
states~\cite{SorensenNATURE2001,WinelandPRA1994} have been
generated~\cite{EsteveNATURE2008,GrossNATURE2010,RiedelNATURE2010,ChapmanNATPHYS2012,LueckeSCIENCE2011,BerradaNATCOMM2013}
and a rich class of entangled non-Gaussian states is predicted to be obtainable~\cite{PezzePRL2009} including maximally entangled Schr\"odinger cat 
states~\cite{MicheliPRA2003,CiracPRA1998}. The production of these fragile states in large systems remains a challenge and efficient methods for
characterization are necessary because full state reconstruction becomes intractable.
Here, we generate a class of non-Gaussian many-particle entangled states and reveal their quantum properties by studying the distinguishability of experimental probability distributions.

\begin{figure}[b!]
\hypertarget{Fig1}{}
\includegraphics[width = 6.1cm]{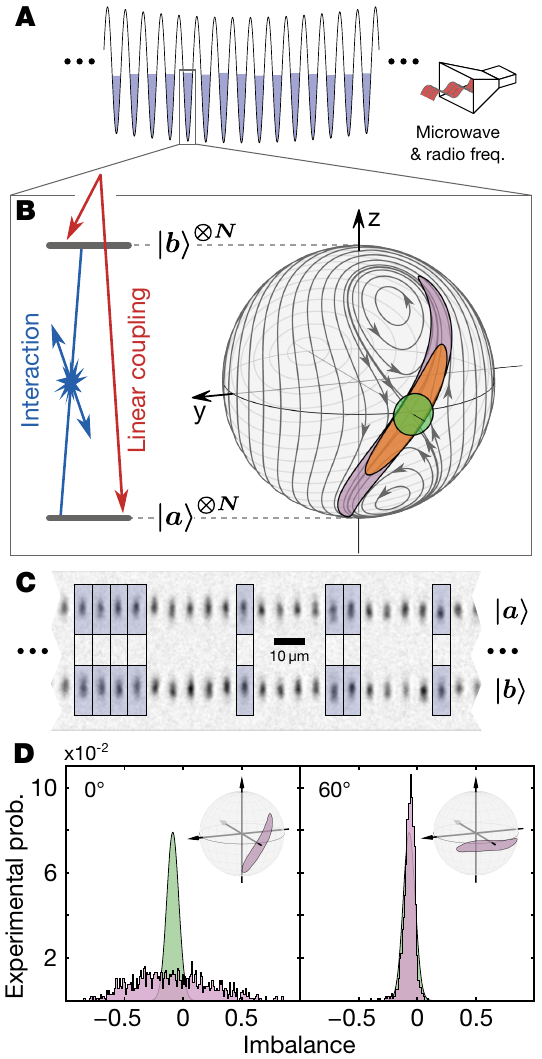}
\vspace{-0.3cm}
\caption{
\textbf{Preparation and detection of non-Gaussian entangled states.} 
\textbf{(A) }
Array of Bose-Einstein condensates in an optical lattice potential addressed by microwave and radio frequency fields.
\textbf{(B) }
The interplay of nonlinear interaction (blue) and weak Rabi coupling (red) between the internal states $\vert a \rangle$ and $\vert b\rangle$ results in an unstable fixed point in the classical phase space.
The state of the system is visualized on a generalized Bloch sphere with radius $J=N/2$. 
Gray lines indicate trajectories of the mean-field equations of motion~\cite{SuppInfo}.
The initial coherent spin state (green) ideally evolves into a squeezed state (orange) followed by non-Gaussian states at later evolution times (violet).
Edges of shaded areas are contours of the Husimi distribution for $N=380$ at $1/\mathrm{e}^2$ of its maximum.
\textbf{(C) }
Experimental absorption picture, showing the site- and state-resolved optical lattice after a Stern-Gerlach separation. 
Shaded boxes indicate the sites with a total atom number in the range $380\pm15$, which are selected for further analysis. 
\textbf{(D) }
Example histograms of the imbalance $z=2J_z/N$ after nonlinear evolution of 25\,ms and final rotation (angles indicated in the panels) compared with the ideal coherent spin state of identical $N$ (green Gaussian).
}
\end{figure}

\begin{figure*}[htp!]
\hypertarget{Fig2}{}
\includegraphics{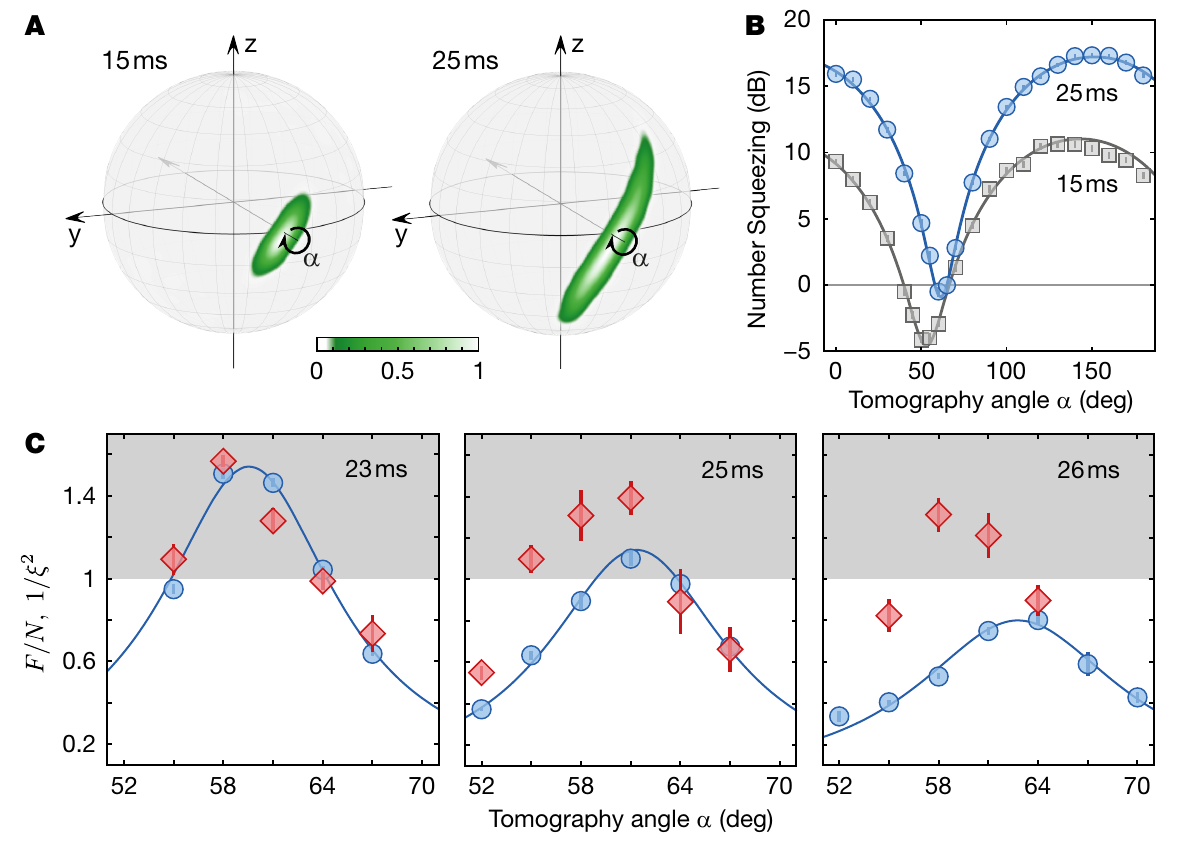}
\vspace{-0.1cm}
\caption{
\textbf{Entangled state characterization.}
 \textbf{(A) }
Tomographic reconstruction of the experimental state after evolution times of 15\,ms and 25\,ms.
The Husimi projections (scaled amplitude is brightness coded~\cite{SuppInfo}) confirm the creation 
of an elongated state and subsequent distortion of the Gaussian shape.
 \textbf{(B) }
Variance analysis of the particle number difference reveals maximum spin squeezing of 
$-4.5\pm0.2$\,dB for 15\,ms and $-0.2\pm0.3$\,dB for 25\,ms.
 \textbf{(C) }
Comparison of the normalized Fisher information $F/N$ (red diamonds) for $N=380 \pm 15$ atoms and the inverted spin squeezing 
factor $1/\xi^2$ (blue circles).
The gray shaded area is only accessible for non-separable (entangled) states.
For 26\,ms spin squeezing cannot identify the entanglement which is detected by the Fisher information. 
Lines are sinusoidal fits, error bars of the Fisher information represent the $68\%$ confidence interval of the Hellinger distance method.}
\end{figure*}

A measure of the distinguishability with respect to small phase changes of the state is provided by the Fisher information 
$F$~\cite{Fisher1925}. It is related to the highest attainable interferometric phase sensitivity by the Cramer-Rao bound 
$\Delta \theta_{\rm CR} = 1/\sqrt{F}$~\cite{HelstromBOOK}.
This limit follows from general statistical arguments for a measurement device with fluctuating output~\cite{SuppInfo}. The Fisher information is limited by quantum fluctuations of the input state as well as the performance of the device.
Even in the absence of technical noise, the Fisher information of a classical input state is $F \le N$ because of the intrinsic granularity of $N$ independent particles which translates into the shot-noise limit $\Delta \theta \geq 1/\sqrt{N}$ for phase estimation.
This classical bound can be surpassed with a reduction of the input fluctuations by introducing entanglement between the $N$
particles~\cite{SorensenNATURE2001}.
These states, known as squeezed states, are fully characterized by mean and variance of the observable and already employed in precision 
measurements~\cite{SchnabelNatComm2010,SewellPRL2012,OckeloenPRL2013}.
In contrast, non-Gaussian quantum states can have increased fluctuations of the observable but nevertheless allow surpassing shot-noise limited performance.
A textbook example is the Schr\"odinger cat state characterized by macroscopic fluctuations but achieving the best interferometric performance allowed by quantum mechanics, i.e. at the fundamental Heisenberg limit
$F=N^2$~\cite{GiovannettiPRL2006}.
In general, the class of states that are entangled and useful for sub shot-noise phase estimation is identified by the Fisher information criterion
$F>N$~\cite{PezzePRL2009}.
Exploiting these resources requires probabilistic methods for phase estimation such as maximum likelihood or Bayesian 
analysis~\cite{KrischekPRL2011} which go beyond standard evaluation of averages.

A paradigm physical process that exhibits the transition from Gaussian to non-Gaussian states is the time evolution 
of a quantum state initially prepared at an unstable classical fixed point.
Our experimental system is an array of interacting binary Bose-Einstein condensates of $^{87}$Rb with additional linear coupling of the two internal states (see \hyperlink{Fig1}{Fig.\,1A--C}), which allows for the controlled realization of such unstable fixed point 
dynamics~\cite{ZiboldPRL2010}. The linear coupling, realized with microwave and radio frequency magnetic fields, also permits precise rotations for initial state preparation and final state manipulation before read-out (see \hyperlink{Fig1}{Fig.\,1B,D}).
For a coherent state initially centered on the unstable fixed point, the quantum dynamics leads to spin squeezed states for short evolution times that subsequently transform into non-Gaussian 
states on an experimentally feasible time
scale~\cite{MicheliPRA2003} (see \hyperlink{Fig1}{Fig.\,1B}).

For the characterization of non-Gaussian states, higher moments, or even the full probability distributions, have to be accessed experimentally. For this, the 
setup~\cite{GrossNATURE2010} has been extended to realize up to 35 individual condensates in a single experiment, which permits the acquisition of sufficient statistics in a narrow window of $\pm15$ for the final atom number. The populations of the atomic states $\left|a\right\rangle = \left|F=1,m_F=1\right\rangle$ and $\left|b\right\rangle = \left | F=2,m_F=-1\right\rangle$ are destructively detected for each individual condensate by state-selective absorption imaging with high spatial resolution (\hyperlink{Fig1}{Fig.\,1C})~\cite{Muessel2013}. By repeating the experiment (typically many thousands of times), we measure the experimental probability distributions of the population imbalance $z=(N_b-N_a)/N$ along defined directions by applying the corresponding spin rotation before detection. The analysis window for $N=N_b+N_a$ is adjusted according to the independently determined time scale of atom loss~\cite{SuppInfo} to follow the time evolution starting with $\langle N\rangle=470$. \hyperlink{Fig1}{Figure 1D} shows examples of observed distributions for two different orientations and an evolution time of 25\,ms; the distributions are consistent with the theoretically expected structure of the state (see \hyperlink{Fig1}{Fig.\,1B}).

Detailed insight can be gained by repeating this measurement for various angles (here in steps of 10 degrees) allowing for the maximum-likelihood reconstruction of the density matrix in the symmetric subspace~\cite{SuppInfo}. \hyperlink{Fig2}{Figure 2A} shows the tomography results obtained from 32,500 experimental realizations confirming qualitatively the expected behavior -- at short evolution times the state has a squeezed shape whereas at later times the characteristic bending dynamics appears as expected from the presence of the two stable fixed points above and below the equator. 

Analyzing the variance of $z$ for the same data as a function of the tomography angle (\hyperlink{Fig2}{Fig.\,2B}) shows that the time evolution leads to suppressed fluctuations at 15\,ms. Extracting the spin squeezing parameter $\xi^2$~\cite{SuppInfo} we find the minimum $\xi_{\rm min}^2 = -4.5\pm0.2$\,dB below the standard quantum limit which demonstrates
entanglement~\cite{SorensenNATURE2001}. We note that, for all results reported here, the photon shot-noise of the absorption imaging of $\pm 4$ atoms is not subtracted. For longer time evolution the bending dynamics leads to increased fluctuations in all directions i.e. tomography angles.  After 25\,ms spin squeezing is lost and we find $\xi_{\rm min}^2 = -0.2\pm0.3$\,dB.
However as shown in \hyperlink{Fig2}{Fig.\,2C}, experimental extraction of the Fisher information (detailed below) reveals that useful entanglement is still present although spin squeezing is vanishing, i.e.
$F/N \ge 1/\xi^2$~\cite{PezzePRL2009}. At 26\,ms, spin squeezing is completely lost whereas the Fisher information still indicates the presence of quantum resources ($F/N>1$). In the Gaussian regime up to 23\,ms, we observe that Fisher information and the inverse spin squeezing agree as expected $F/N \approx 1/\xi^2$ as these states are fully characterized by their variance.

\begin{figure}[htp!]
\hypertarget{Fig3}{}
\includegraphics{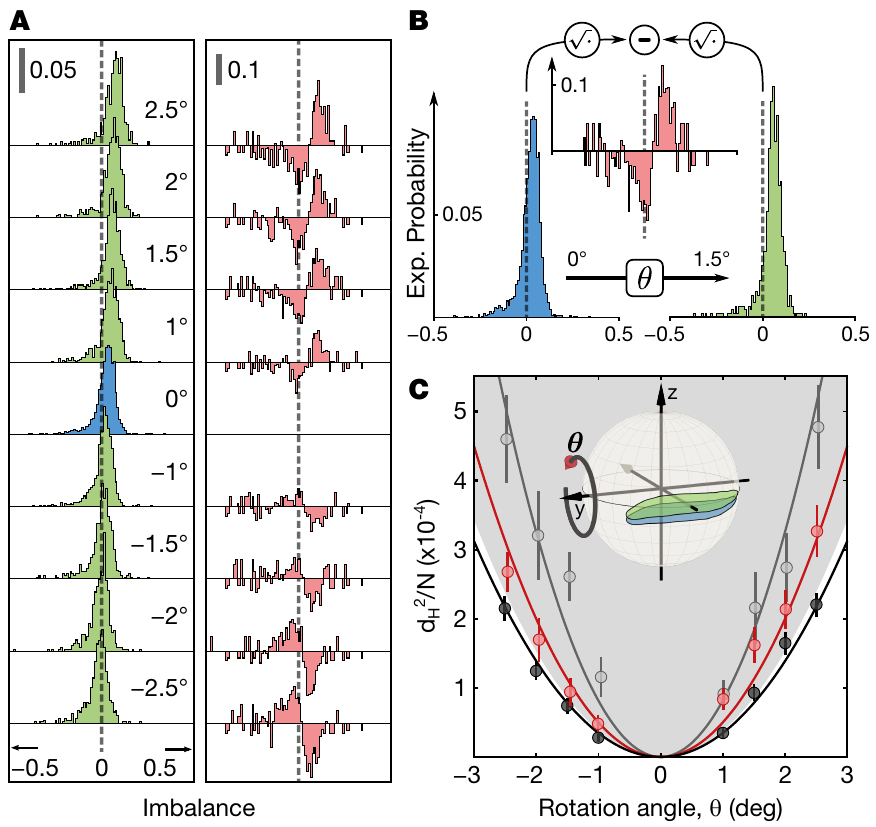}
\caption{
\textbf{Experimental extraction of the Fisher information.}
 \textbf{(A) }
Examples of experimental histograms of $z$ (left panel) after 26\,ms of evolution time 
obtained for the tomography angle $\alpha=58^{\circ}$ and additional small rotation angles $\theta$ about the y-axis (see inset of \textbf{C}).
The blue histogram is used as a reference for the following analysis. 
The right panel shows the square-root-differences between the respective histogram and the reference (see panel B). Gray bars indicate the vertical scales (see axes in panel B).
 \textbf{(B) }
Procedure for extraction of the squared Hellinger distance (example of $\theta=1.5^{\circ}$).
Probability amplitudes $\sqrt{P_z}$ are subtracted for each bin and these differences squared and summed to obtain the squared Hellinger distance.
 \textbf{(C) }
Squared Hellinger distances for three different evolution times. 
A reference measurement (black) with the initial coherent spin state (0\,ms evolution) lies slightly outside the non-classical region (gray shaded area). 
The spin squeezed state (gray, 15\,ms evolution) surpasses the classical limit. After 26\,ms (red points, histograms shown in \textbf{A}) the state has a non-Gaussian shape, is not spin squeezed but still performs beyond the standard quantum limit. 
Error bars indicate the statistical $68\%$ confidence interval obtained by a resampling procedure~\cite{SuppInfo}. The curvature of the quadratic fit is proportional to $F/N$.
}
\end{figure}

Our method for extraction of the Fisher information circumvents the experimentally intractable full reconstruction of the density matrix and is based on a specific set of experimental probability distributions $P_z(\theta)$  after small rotations $\theta$ of the quantum state. As an example, in \hyperlink{Fig3}{Fig.\,3A} we show distributions for the state created at 26\,ms and the optimal tomography angle of 58 degrees (see \hyperlink{Fig2}{Fig.\,2C}) after small rotations about the $y$ axis (see \hyperlink{Fig3}{Fig.\,3C} inset), which feature a  pronounced peak and long tails characteristic for the bent state.
The main effect of the small rotation is a continuous shift of the distribution towards increasing imbalance for larger angles $\theta$.

The analysis for extraction of the Fisher information builds on the statistical 
distance~\cite{WoottersPRD1981,BraunsteinPRL1994}
of these distribution functions. We use a Euclidean distance in the space of probability amplitudes $\sqrt{P_z}$ known as Hellinger distance~\cite{SuppInfo}, defined as 
\begin{equation}
d_\mathrm{H}^2(\theta) = \frac{1}{2} \sum_{z} \left( \sqrt{P_z(\theta)} - \sqrt{P_z(0)} \right)^2.
\end{equation}
According to the definition, we take the square root of the experimental probability distributions at finite $\theta$ and $\theta=0$ and calculate the difference for each bin (red histograms in \hyperlink{Fig3}{Fig.\,3A,B}). Summing the squares of the differences provides the squared Hellinger distance $d_\mathrm{H}^2(\theta)$. \hyperlink{Fig3}{Figure 3C} shows $d_\mathrm{H}^2(\theta)$ as a function of $\theta$ for three different evolution times and the respective optimal tomography angles. A resampling method~\cite{SuppInfo} is used to extract error bars and reduce the statistical bias.
The observed quadratic behavior is expected from the Taylor expansion 
\begin{equation}
d_\mathrm{H}^2(\theta) = \frac{F}{8}\,\theta^2 + \mathcal{O}(\theta^3),
\end{equation}
which reveals the close connection between Hellinger distance and Fisher information; this relationship is used to extract $F$ from the curvature of $d_\mathrm{H}^2(\theta)$.
The gray shaded area in \hyperlink{Fig3}{Fig.\,3C} indicates the region that is not accessible to separable states; for them, the Fisher information is limited to $F/N \le 1$, resulting in a $d_\mathrm{H}^2$ curvature smaller than $N/8$. For the initially prepared state we find a Fisher information of $F/N=0.91\pm 0.04<1$, which is expected for a separable state. For a subsequent evolution of 26\,ms the measured Hellinger distances lie in the regime of non-separability. This reveals entanglement in a regime where no spin squeezing is present. For the intermediate evolution time of 15\,ms we extract a Fisher information $F/N=2.2\pm 0.2$, which confirms entanglement in the Gaussian spin squeezed state. For obtaining the systematic study of \hyperlink{Fig2}{Fig.\,2C}, this procedure is performed at a given evolution time for different tomography angles.
The reported values for the Fisher information are limited by experimental imperfections, detection noise and atom loss which especially affect the fragile non-Gaussian states. For the ideal time evolution, monotonically increasing Fisher information is expected, whereas the available spin squeezing is limited to $1/\xi^2\approx 18$ (-12.6\,dB). The ideal theoretical model prediction is $F/N\approx 90$ (-19.5\,dB) for the evolution time when spin squeezing vanishes~\cite{SuppInfo}.

There is a direct connection between Fisher information and sensitivity in parameter estimation. In an interferometric context, high sensitivity, indicated by a large value of $F$, means fast change of the output distribution with the phase $\theta$, i.e. high statistical speed $\partial d_\mathrm{H} / \partial\theta = \sqrt{F/8}$ with respect to the parameter change. For the quantum state at 25\,ms the enhanced Fisher information reveals quantum resources beyond the standard quantum limit in a range of tomography angles. No squeezing is detected for the optimum at 58 degrees, which implies that these resources can only be exploited with the knowledge of more details of the distribution functions. In \hyperlink{Fig4}{Fig.\,4A} we show explicitly through mean and variance analysis that averaging of the observable $z$ does not surpass shot-noise limited performance for rotations about the $y$ axis (corresponding for example to a phase shift inside a Ramsey interferometer~\cite{SuppInfo}).

\begin{figure}[t!]
\hypertarget{Fig4}{}
\includegraphics[width = 6.5cm]{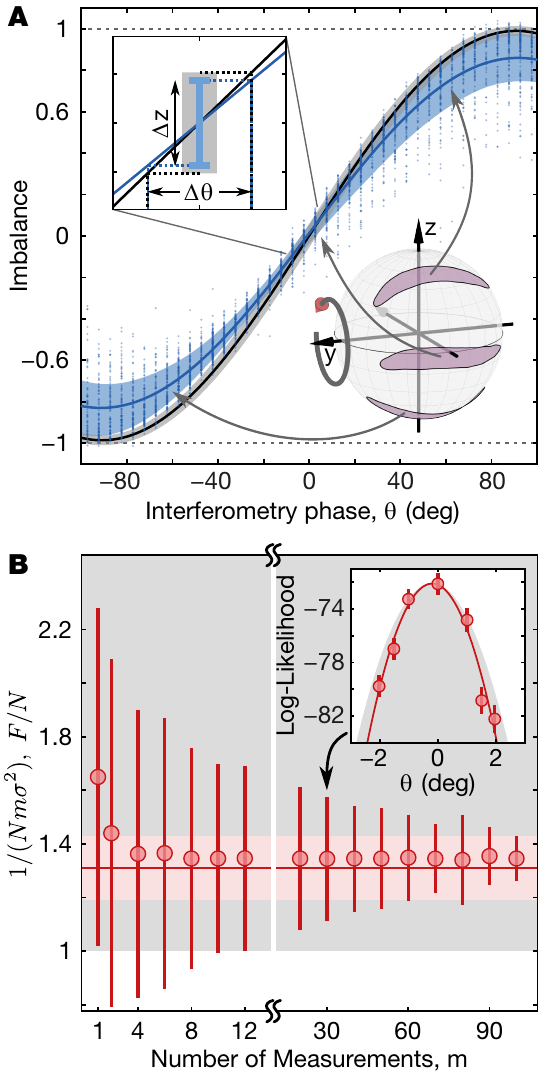}
\caption{
\textbf{Quantum enhanced phase sensitivity in the absence of spin squeezing.} 
\textbf{(A) } A rotation on the Bloch sphere by the angle $\theta$ (right inset) is formally equivalent to the action of a 
Ramsey interferometer with relative phase shift $\theta$ (see \protect\hyperlink{FigS5}{Fig.\,S5}).
Interference fringe of z (blue line with $68\%$ confidence interval) with the state after 25\,ms evolution and subsequent tomography rotation of $61$ degrees; reference measurement with a coherent spin state (black line with gray confidence region).
The phase sensitivity $\Delta \theta$ deduced from standard error propagation via the slope of the interferometer fringe (left inset) does not surpass the standard quantum limit.
 \textbf{(B) }
Bayesian analysis with the tomography angle for maximal Fisher information (58 degrees).
For each sequence of length $m$ a quadratic fit to $\log \mathcal{L}$ is used to extract the Gaussian variance $\sigma^2$ of $\mathcal{L}$ which corresponds to the phase sensitivity.
$1/(N m \sigma^2)$ as a function of $m$ shows fast convergence to the extracted $F/N$ using the Hellinger distance method (red line with $68\%$ confidence region). 
The error bars indicate the standard deviation of a single sequence. Gray shaded regions are only accessible with entanglement in the system.
}
\end{figure}

However, the resource can be harnessed with model independent Bayesian estimation using the experimental probability distributions $P_z(\theta)$.
For this, we use an independent data set taken with the setting ($\alpha=58^{\circ}$, $\theta_0=0^{\circ}$) and 25\,ms of evolution time, which we divide into sequences $\{z_1, \dots, z_m\}$ each containing $m$ realizations. To obtain realistic measurement conditions, we discard the previous knowledge on the true value of the phase $\theta_0$ (Bayesian estimation with flat prior~\cite{SuppInfo}).
For each distribution $P_{z}(\theta_i)$ we calculate the likelihood
\begin{equation}
\mathcal{L}(\theta_i) = \prod_{j=1}^m P_{z_j}(\theta_i),
\end{equation}
which corresponds to the conditional probability to obtain the sequence $\{z_1, \dots, z_m\}$ if the phase setting had been $\theta_i$.
For sufficiently large $m$ we expect a Gaussian distribution $\mathcal{L}(\theta) \propto \exp(-(\theta-\theta_c)^2/2\sigma^2)$ centered at $\theta_c$ with the Bayesian phase uncertainty $\sigma$~\cite{SuppInfo}. Thus, $\sigma$ can be extracted from a quadratic fit to $\log\mathcal{L}(\theta)$ (see \hyperlink{Fig4}{Fig.\,4B} inset).
We find a fast convergence of $\sigma^2$ to the expected value $1/m F$ (Cramer-Rao bound for $m$ measurements) already for $m \gtrsim 2$ (\hyperlink{Fig4}{Fig.\,4B}). This explicitly shows phase uncertainty below the standard quantum limit in agreement with the Fisher information obtained from the Hellinger distance method described above.

In conclusion, we have developed a method based on the statistical distance of experimental probability distributions to extract the Fisher information, which was previously unattainable in systems of large particle number. We demonstrate this on a novel class of collective states in binary Bose-Einstein condensates generated in the vicinity of a classically unstable point. 
The experimental value of the Fisher information serves to verify entanglement in the absence of spin squeezing and quantifies the quantum resource for improved phase estimation. 
We confirm this by characterizing the sensitivity of a Ramsey interferometer and find enhanced performance in agreement with the extracted Fisher information. The presented method does not depend on the special shape of the probability distributions and is not limited 
to small particle numbers. It is therefore broadly applicable to the efficient characterization of 
highly entangled states, relevant for further improvement of atom 
interferometers~\cite{AppelPNAS2009,GrossNATURE2010,LerouxPRL2010,LueckeSCIENCE2011,ChenPRL2011,SewellPRL2012,BerradaNATCOMM2013,OckeloenPRL2013} 
toward the ultimate Heisenberg 
limit~\cite{GiovannettiPRL2006}.
More generally, it can be applied to any phenomenon
characterizable by the distinguishability of quantum states, as in quantum phase transitions~\cite{Zanardi}, quantum Zeno 
dynamics~\cite{SmerziPRL2012} and quantum information 
protocols~\cite{NielsenBOOK}.

\textbf{Acknowledgements}
We thank J. Tomkovi\v{c}, E. Nicklas and I. Stroescu for technical help and discussions.
This work was supported by 
the Forschergruppe FOR760, 
the Deutsche Forschungsgemeinschaft, 
the Heidelberg Center for Quantum Dynamics and 
the European Commission small or medium-scale focused research project QIBEC (Quantum Interferometry with Bose-Einstein condensates, Contract Nr. 284584).
W.M. acknowledges support by the Studienstiftung des deutschen Volkes.
D.B.H. acknowledges support from the Alexander von Humboldt foundation.
L.P. acknowledges financial support by MIUR through FIRB Project No. RBFR08H058.
QSTAR is the MPQ, LENS, IIT, UniFi Joint Center for Quantum Science and
Technology in Arcetri.

\textbf{Published in} Science {\bf 345}, 424-427 (2014)\\
doi: \href{http://dx.doi.org/10.1126/science.1250147}{10.1126/science.1250147}\\
PDF also available via \\
\href{http://www.matterwave.de/publications}{http://www.matterwave.de/publications}
%
%
%
%

\stepcounter{myequationcnt}
\renewcommand{\theequation}{S\arabic{equation}}
\stepcounter{myfigurecnt}
\renewcommand{\thefigure}{S\arabic{figure}}
\hyphenation{Fesh-bach}
%
\section{Supplementary Materials}
\hypertarget{HyperSuppInfo}{}
\vspace{-0.3cm}
\subsection*{Experimental system}
\vspace{-0.2cm}
A Bose-Einstein condensate of $^{87}$Rb atoms is loaded into a one-dimensional optical lattice superimposed with a shallow harmonic trap.
The trap frequencies are $660\,\text{Hz}$ in lattice direction and $260\,\text{Hz}$ in radial direction. In this tight confinement regime, 
the spin healing length is on the order of the size of the on-site wavefunction such that the single-mode approximation is applicable.
The large lattice spacing ($5.5\,\text{\textmu m}$) and the high inter-well potential barrier ensure that tunneling is negligible on the experimental timescale.
The condensate is distributed over up to 35 independent lattice sites with occupation numbers ranging from $100$ to $600$ atoms.
In the reported experiments we address the two states $\vert a \rangle=\vert F=1,m_F=1\rangle$ and $\vert b \rangle=\vert F=2,m_F=-1\rangle$ 
of the ground state hyperfine manifold by application of microwave and phase controlled radio frequency radiation, which drive a magnetic two-photon transition.
Nonlinear interaction between the condensate atoms is enhanced at a magnetic field of $9.12\,\text{G}$ in the vicinity of an inter-species Feshbach resonance.
An active stabilization, including a feed-forward of the $50\,\text{Hz}$ mains frequency, reduces the shot-to-shot fluctuations 
of the offset field below $30\,\text{\textmu G}$.

The main physical limitation of the experimental system is the effect of particle losses, which leads to noise contributions as well as a mean change in the interaction 
dependent parameters (detailed below) during the evolution time. The most limiting loss channel is dipole relaxation of the $F=2$ manifold, by which in every event two
atoms of $|b\rangle$ get lost from the trap~\cite{TojoPRA2009}. The corresponding $1/\mathrm{e}$ decay time of pure $|2,-1\rangle$ is $\sim 200\,\text{ms}$. Additionally, the enhancement of inelastic scattering and three-body recombination caused by the 
closeby Feshbach resonance leads to loss of $|a\rangle$ and $|b\rangle$, which is symmetric on average. The combined loss leads to a decay time of $\sim 110\,\text{ms}$ for the total number of atoms.

After the experimental sequence, a resonant $\pi$-pulse transfers the population of $\vert b \rangle$ to $\vert F=1,m_F=-1\rangle$ to stop further dipole relaxation loss in $F=2$. This allows for the controlled ramp-down of the magnetic field to $\sim 1\,\text{G}$ for absorption imaging with a precision of $\pm4$ atoms for the individual atom numbers $N_a$ and $N_b$ on each lattice site~\cite{Muessel2013}. A short time of flight of $1.2\,\text{ms}$ after Stern-Gerlach separation reduces the optical density to achieve optimal conditions for imaging.
\vspace{-0.3cm}
\subsection*{Theoretical description}
\vspace{-0.2cm}
\begin{figure}[t!]
\includegraphics[scale=1]{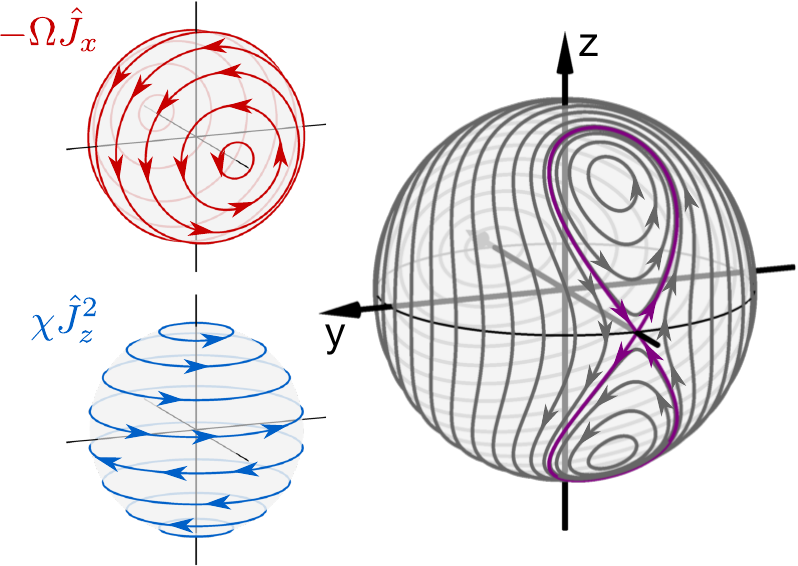}
\vspace{-0.3cm}
\caption{{\bf Schematic illustration of the Hamiltonian.} The two main contributions are linear Rabi coupling with Rabi frequency $\Omega$ (red) and nonlinear interaction (one-axis twisting) with strength $\chi$ (blue). The effect of each part is visualized by several mean-field trajectories for the case of $\chi=0$ and $\Omega=0$ respectively. Arrows indicate the direction of movement. In the regime of $\Lambda=N\chi/\Omega>1$ the combination leads to a mean field phase space with three stable and one unstable fixed point (big sphere with $\Lambda=1.5$). A separatrix (violet line) divides the phase space into three regions with macroscopically different temporal behavior.}
\label{SuppFig1}
\end{figure}
Our system of $N$ two-level atoms in each individual condensate is conveniently described by the collective spin operator
$\vec{J} = \sum_{i=1}^N \vec{\sigma}_i/2$ with components ${\hat{J}_x = (b^\dag a + a^\dag b)/2}$, ${\hat{J}_y =  (b^\dag a - a^\dag b)/2\i}$ and ${\hat{J}_z =  (b^\dag b-a^\dag a)/2}$, 
where $\vec{\sigma}_i$ is the Pauli vector of the $i$th particle and $a^\dag$ and $b^\dag$ are the respective creation operators associated with the 
two modes.
$\hat{J}_z=(\hat{N}_b-\hat{N}_a)/2$ is half the occupation number difference in the two modes while
$\hat{J}_x$ and $\hat{J}_y$ are the corresponding coherences.
The quantum dynamics of the $N$ particles on each lattice site is described by the two-mode Josephson Hamiltonian 
\begin{equation}
\hat{H} = \chi(N) \hat{J}_z^2 - \Omega \hat{J}_x + \delta(N) \hat{J}_z,
\end{equation}
which is a special case of the more general Lipkin-Meshkov-Glick Hamiltonian. The parameters $\chi$, $\Omega$ and $\delta$ are the nonlinearity due to 
atom-atom interaction, the linear coupling strength and the detuning, respectively.

\begin{figure}[tp!]
\includegraphics[scale=1]{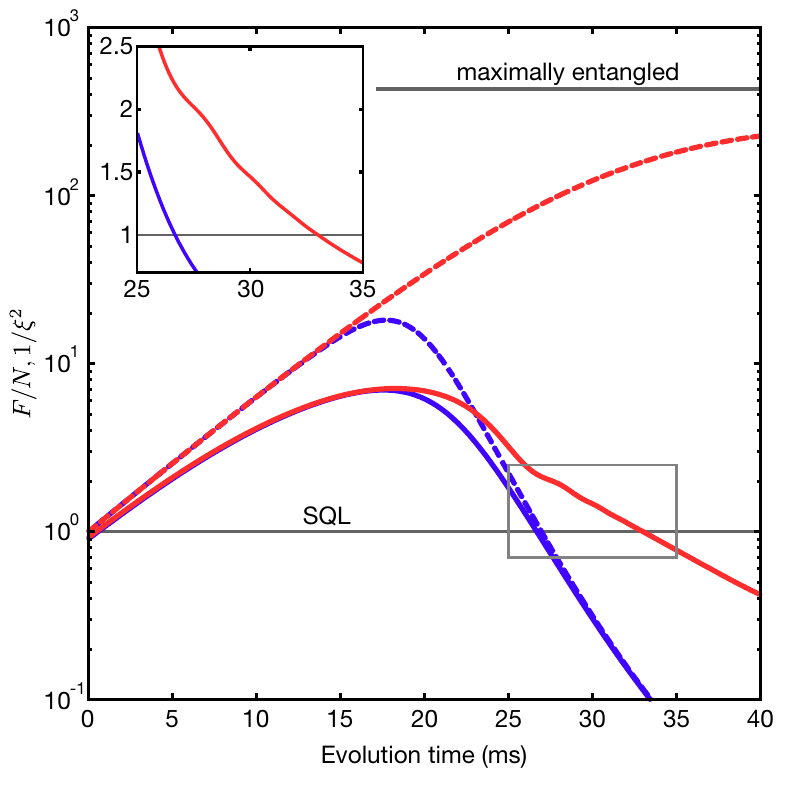}
\vspace{-0.3cm}
\caption{{\bf Time evolution of Fisher information and squeezing}
The dashed lines show the ideal model predictions for constant parameters ($\Lambda=1.5$, $\Omega=2\pi\times20\,\text{Hz}$, $N=430$ and $\delta=0$) obtained by exact diagonalization. Both Fisher information (red) and spin squeezing (blue) are optimized over tomography angle and readout axis.
For short evolution times, Fisher information and the corresponding value $1/\xi^2$ agree. The bending dynamics leads to a decrease of squeezing for later evolution times whereas the Fisher information monotonically increases. Taking into account our finite detection noise ($\pm4$) for the atom number (solid lines) reveals that detection noise is the main limitation for harnessing highly entangled states in an otherwise perfect system. The lower straight line corresponds to the standard quantum limit, the upper straight line is the value $F=N^2$ for a maximally entangled state. The inset shows a zoom into the region indicated by the grey box but scaled linearly.
}
\label{SuppFig5}
\end{figure}

In the ideal case (constant parameters with $N\chi>\Omega$ and $\delta=0$) the early dynamics yield monotonically increasing Fisher information for an initial coherent spin state centered on the classically unstable fixed point (eigenstate of $\hat{J}_x$). In contrast, the maximally attainable spin squeezing is limited
due to the bending dynamics, which leads to an increase of the quantum uncertainty in all spin directions and a reduction of the spin squeezing for later evolution times. Fig.~\ref{SuppFig5} shows numerical simulations of this ideal situation with typical experimental parameters of $N\chi/\Omega=1.5$, $\Omega=2\pi\times20\,\text{Hz}$ and $N=430$, where both Fisher information and spin squeezing have been optimized over rotation and measurement axis for each evolution time.
This shows that the evolution creates non-Gaussian states useful for quantum metrology going beyond the quantum resources of the spin squeezed states accessible with this system.
The Fisher information is found to saturate the Quantum Fisher Information for all evolution times, which is the maximum over all positive operator valued measures (POVM). This shows that the atomic imbalance stays the best observable.
The experimentally observed decrease of the Fisher information as a function of time is a consequence of detection noise, technical imperfections and losses (detailed below).
This basic behaviour is captured by including our detection noise of $\pm 4$ atoms as a Gaussian convolution of the theoretical probability distributions (Fig.~\ref{SuppFig5}, solid lines). The experimentally extracted Fisher information provides a lower bound for the quantum resources as it also includes all technical limitations.
Since the quantum resources of the non-Gaussian states increasingly manifest themselves in substructures on small scales, they are especially affected by imperfect detection, which cannot resolve these features.

In the experimental system $\delta$~and $\chi$ are both a function of the atom number~\cite{MicheliPRA2003}, caused by 
the dependence of the shape of the BEC mean-field order parameter and the mean atom density on $N$. We find experimentally in good approximation $\delta(N) \approx 2\pi \times \left(\delta_0 - \delta_N \sqrt{N} \right) $ with $\delta_0 = 16.3$\,Hz and $\delta_N=0.68\,\text{Hz}/\sqrt{\text{atom}}$. The nonlinearity $\chi(N)$ is $\propto 1/\sqrt{N}$ for our range of atom numbers and $\chi_0\sim 2\pi\times 0.064\,\text{Hz}$ for an initial atom number of $N_0=470$. Besides noise effects, atom loss during the time evolution thus leads to a time dependence of these parameters.

\begin{figure}[h!]
\includegraphics[scale=0.7]{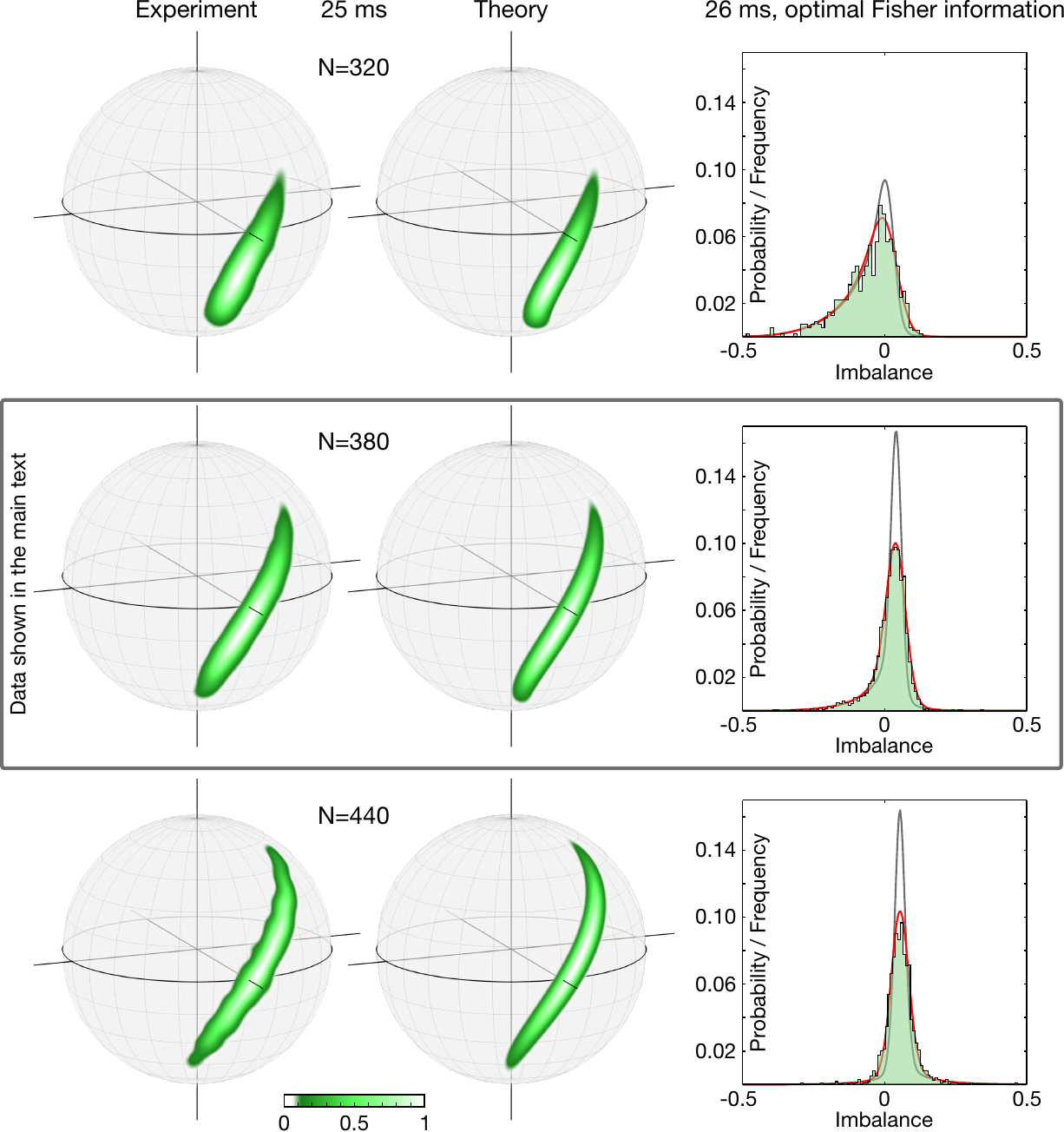}
\vspace{-0.3cm}
\caption{{\bf Atom number dependence of the Hamiltonian.} Experimental Husimi distributions of tomographically reconstructed density matrices (left panel) for $N=320$, 380 and 440 after $25\,\text{ms}$ of evolution time reveal the strong atom number dependence of the shape of the final state (a change of 60 atoms corresponds to a change of $\approx 1\,\text{Hz}$ in detuning for the final atom number). The experimental distributions are in good agreement with numerical simulations (middle panel) including the atom number dependence of the parameters, which also dynamically change due to atom loss.
Finite detuning during state generation also shows up in the experimental probability distributions for $\alpha=58$\textdegree\;(maximal Fisher information) as depicted in the right panel for an evolution time of $26\,\text{ms}$. For the corresponding result from the theoretical model detection noise is included as a Gaussian convolution with $\sigma = 6$ atoms (grey curves). 
Assuming additional Gaussian noise due to atom loss (total $\sigma = 12$, see text), the distributions are in very good agreement with the numerical simulations (red curves).}
\label{SuppFig4}
\end{figure}

Due to these relations, the shape of the final state strongly depends on atom number, as depicted in Fig.~\ref{SuppFig4}. The Husimi distributions for the final atom numbers $N=320$, $380$ and $440$, obtained  from the experimental tomographic reconstruction after an evolution time of $25\,\text{ms}$, reveal a pronounced change in shape of  the state with $N$.  This is consistent with a numerical simulation (Fig.~\ref{SuppFig4} middle panel) taking into account the nonlinearity $\chi=\chi_0\sqrt{N_0/N(t)}$ and detuning of $\delta(N(t))$. This includes the atom number dependence of the parameters during the time evolution, assuming $N(t) = N_0 e^{-t/\tau}$ with the experimentally determined decay time $\tau=110\,\text{ms}$ due to atom loss.
All fast pulses (with $-\Omega (\cos\phi\hat{J}_x + \sin\phi\hat{J}_y)$ in the Hamiltonian), including the spin-echo pulse in the middle of the time evolution, are modeled in the presence of nonlinearity and with a phase offset of $\delta\phi=3$\textdegree , which was employed in the experiment to compensate for nonlinearity during the initial preparation pulse. 
The experimentally observed shape is well explained by the assumed model. It is mainly the finite detuning which leads to the asymmetry of the final state.
This is reflected in the Husimi projections and also shows up as an asymmetry of the experimental probability distributions for the tomography angle $\alpha=58$\textdegree\;yielding maximal Fisher information (right column of Fig.~\ref{SuppFig4} for an evolution time of $26\,\text{ms}$). 
For a comparison with the model, detection noise of the absorption imaging is taken into account by a convolution of the final distributions with a Gaussian of width $\sigma_{\text{det}} = 6$\,atoms for $N_b-N_a$ (grey curves).
Furthermore, the effect of particle loss of $\sim 100\;\text{atoms}$ up to this evolution time leads to additional fluctuations which we estimate as $\sigma_{\text{loss}} \approx 10$, yielding a total width of $\sigma_{\text{tot}} \approx 12$ atoms. With this convolution, the numerical simulations (red curves) are consistent with the observed experimental probability distributions.
\vspace{-0.3cm}
\subsection*{Classical phase space}
\vspace{-0.2cm}
A useful insight to the dynamics is offered by the mean field picture which is exactly valid for ${N\rightarrow\infty}$ and obtained by replacing the
quantum mechanical operators by their mean values $(\langle \hat{J}_x\rangle ,\langle \hat{J}_y \rangle,\langle \hat{J}_z \rangle) = (N/2)(\sqrt{1-z^2}\cos\phi,\sqrt{1-z^2}\sin\phi,z)$,
 where the imbalance $z = (N_b-N_a)/N$ is the normalized population difference and $\phi$ is the relative phase between the two internal states.
In this limit, the dynamical evolution can be gathered in classical equations of motion of $z$ and 
its conjugate variable $\phi$~\cite{SmerziPRL,ZiboldPRL2010}. The Hamiltonian becomes 
\begin{equation}
H = \frac{N\Omega}{2}\left( \frac{\Lambda}{2} z^2 - \sqrt{1-z^2} \cos \phi + \frac{\delta}{\Omega} z \right),
\end{equation} 
which is formally equivalent to a non-rigid pendulum, where $\Lambda = N \chi/\Omega$.
The equipotential lines of this classical Hamiltonian are the classical trajectories. These are shown in Fig.~\ref{SuppFig1} for 
the three cases $\Omega \gg N\chi$ (red trajectories), $\Omega = 0$ (blue trajectories) and $\Omega = N\chi/1.5$ (gray trajectories), 
corresponding to the three cases of dominating Rabi coupling, 
dominating nonlinear evolution and the Josephson regime of weak Rabi coupling, respectively. For $\Omega\ne0$ the topology of the phase space 
only depends on the parameter $\Lambda$. For $\Lambda > 1$ and $\delta=0$ the phase space $(z,\phi)$ features three stable fixed points $((0,0)\; \text{and}\; (\pm \sqrt{1- 1/\Lambda^2}, \pi))$ and the unstable fixed point $(0,\pi)$ on the negative x-axis. 
An eight-shaped separatrix passing through the unstable fixed point divides the phase space into three regions of macroscopically different
temporal behavior~\cite{ZiboldPRL2010,SmerziPRL}.
\vspace{-0.3cm}
\subsection*{Experimental sequence}
\vspace{-0.2cm}
\begin{figure}[h]
\includegraphics[scale=1]{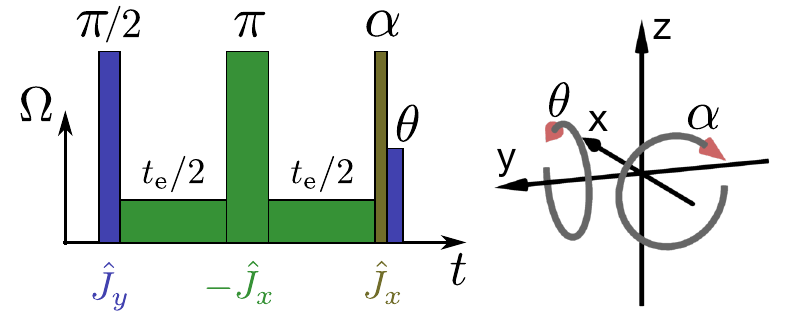}
\vspace{-0.3cm}
\caption{{\bf Experimental sequence for state generation and characterization.} The first $\pi/2$ pulse prepares every atom in an equal superposition of the two internal states $|a\rangle$ and $|b\rangle$, corresponding to the initial coherent spin state. The preparation on the unstable fixed point is accomplished by a nonadiabatic change of the radio frequency phase and attenuation of $\Omega$ for the evolution time $t_e$ (not to scale). A spin echo $\pi$-pulse is applied in the middle of the time evolution to suppress external detuning fluctuations. After state generation, rotation pulses (indicated by their respective angle) are applied for state characterization. The arrows in the coordinate system indicate the corresponding rotation axes.}
\label{SuppFig2}
\end{figure}
The experiment starts with a BEC of $\left|1,-1\right\rangle$ in each potential well. 
After ramping up the magnetic field to $9.12\,\text{G}$, a radio frequency rapid adiabatic passage 
transfers it to $\vert a \rangle^{\otimes N} = \left|1,+1\right\rangle^{\otimes N}$.
In Fig.~\ref{SuppFig2} we show schematically the experimental sequence beginning from this initial state.
A fast $\pi/2$ Rabi pulse and subsequent non-adiabatic change of the driving phase by $3\pi/2$
prepares each condensate in the coherent spin state $(\vert b \rangle - \vert a \rangle)^{\otimes N}/2^{N/2}$, 
which corresponds to an independent superposition of each atom between the states $\left|a\right\rangle$ and $\left|b\right\rangle$ 
with the mean spin direction pointing towards the unstable fixed point.
The state preparation pulses are sufficiently fast such that nonlinear effects induced by particle-particle interaction are negligible.
The regime $\Lambda > 1$ is addressed by attenuating the Rabi coupling strength
below the nonlinear interaction $N\chi$.
In the middle of the time evolution, a fast $\pi$ spin-echo pulse 
around the negative $x$-axis is applied in order to further reduce the effect of shot-to-shot fluctuations of the 
detuning $\delta$ (caused by 
the finite magnetic field stability).

The Rabi pulses are implemented with microwave and radio frequency 
magnetic fields $200\,\text{kHz}$ red-detuned with respect to the 
$\left|1,1\right\rangle \leftrightarrow \left|2,0\right\rangle$ and $\left|2,-1\right\rangle \leftrightarrow \left|2,0\right\rangle$ transition, respectively. 
The resulting two-photon Rabi frequency for preparation, $\pi$-pulse and tomography rotation is $\sim 320\,\text{Hz}$ (corresponding to $\Lambda \approx 0.1$) and is calibrated by Rabi flopping. The last $\theta$-rotation is performed with an independently calibrated Rabi frequency of $\sim 160\,\text{Hz}$ to reduce the influence of spurious timing and switching effects.
\begin{figure}[t]
\hypertarget{FigS5}{}
\includegraphics[scale=1]{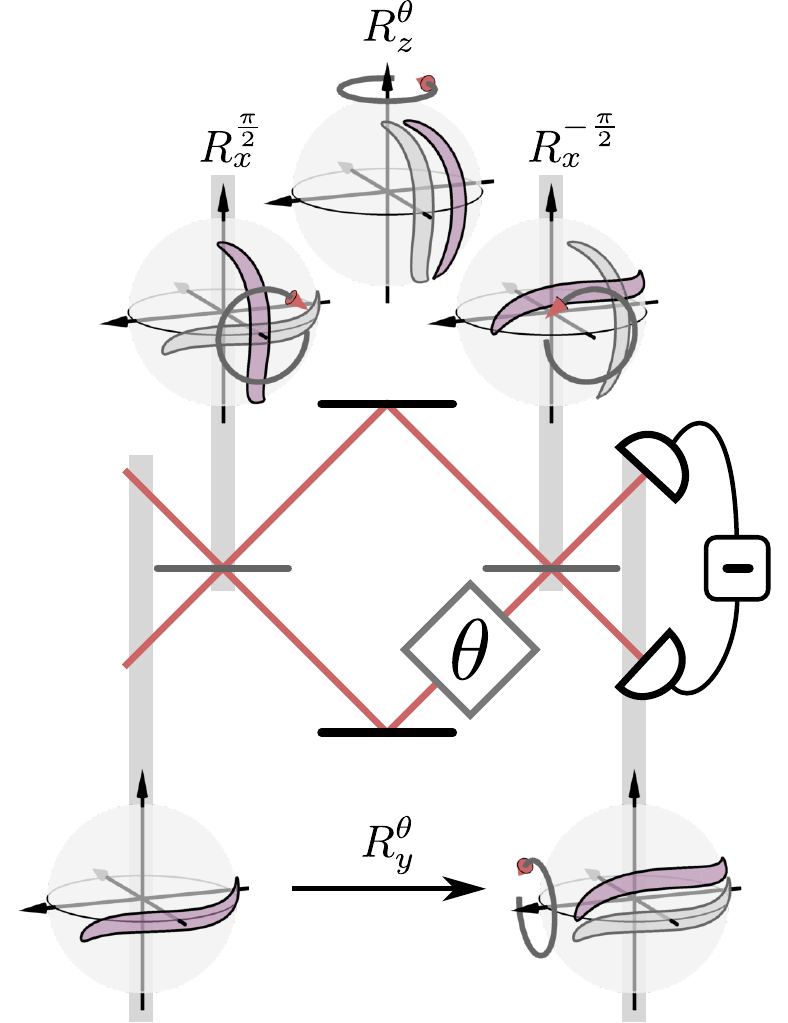}
\vspace{-0.3cm}
\caption{{\bf Linear SU(2) interferometer and rotations on the Bloch sphere.} The Mach-Zehnder interferometer (shown as sketch) is the optical analog of the atomic Ramsey sequence. Balanced beamsplitters ($\pi/2$-pulses) perform rotations $R_x$ by an angle of $\pi/2$ and the relative phase shift inside the interferometer performs a rotation $R_z^\theta$ by the angle $\theta$. $R_{x,y,z}$ are the corresponding rotation matrices of SO(3). The sequence $R_x^{-\pi/2} R_z^{\theta} R_x^{\pi/2}$ is mathematically equivalent to $R_y^{\theta}$. For a characterization of the interferometric sensitivity it is thus sufficient to apply $R_y^\theta$ only.
}
\label{SuppFig3}
\end{figure}
\vspace{-0.3cm}
\subsection*{Spatial homogeneity}
\vspace{-0.2cm}
Special care has to be taken to ensure homogeneity of all applied fields along the one-dimensional lattice 
since we want lattice sites separated by up to $\sim 100\,\text{\textmu m}$ to bear consistent conditions. 
We compensate the gradient of the magnetic offset field by adjusting small permanent magnets near the experimental chamber such 
that a magnetically sensitive microwave Ramsey sequence on the transition $|1,1\rangle \leftrightarrow |2,2\rangle$ no longer shows any observable periodic pattern along the lattice. 
We note, that this measurement can be performed 
with higher accuracy than the on-site magnetic field stability because we have the advantage of the many-well information in every shot. It is only limited by the collisional (mean-field) shift of the two employed levels, which depends 
on the number of atoms on each lattice site. The relevant gradient of the microwave magnetic field can be measured by Rabi flopping on the transition $|1,1\rangle\leftrightarrow|2,0\rangle$ with many cycles ($\sim 50$), which eventually dephases along the lattice. This was minimized by small displacements of the microwave antenna and we find a remaining 
power gradient of $3.2\,\%$ over the whole ensemble which probably stems from the surrounding metal parts.
For the fast rotation pulses, the influence of this remaining gradient on the homogeneity of the Rabi coupling is partly compensated by the gradient of the radio frequency power, which is on the same order but with opposite sign due to the special geometry and position of the coil of the radio frequency antenna. The remaining gradient of the Rabi coupling is $\sim 0.7\%$ over the whole BEC array, by which we can constrain the inhomogeneity of a $\pi/2$-pulse to $\lesssim 0.3$ degrees in the relevant range of atom numbers with a maximum spacing of $20$ wells. This corresponds to $\sim 5\%$ of the width  (2 standard deviations) of a coherent spin state with $400$ atoms.

For the two-photon coupling, the gradients of the off-resonant microwave and radio frequency driving also cause inhomogeneity of the AC Zeeman shifts.
The absolute values are $\sim 120\,\text{Hz}$ (microwave) and $\sim 70\,\text{Hz}$ (radio frequency) which add up to $\sim 190\,\text{Hz}$ total shift of the
two-photon resonance in the case of the maximal power used for the fastest pulses.
In order to reduce the influence of these gradients on the detuning during the time evolution, we distribute the attenuation over the two contributions. The microwave power is reduced by $11\,\text{dB}$ with an attenuator on a fast MW switch with two ports. The remaining $14\,\text{dB}$ attenuation needed to reach $\Lambda \approx 1.5$ is achieved by reducing the power of the radio frequency. In this way, the gradient of the detuning $\delta$ during the time evolution can be reduced to below $2\pi\times0.3\,\text{Hz}$ over the whole ensemble.
\vspace{-0.3cm}
\subsection*{Long term stability and systematics}
\vspace{-0.2cm}
The most critical parameter in the experimental sequence is the detuning $\delta$ during the time evolution. Because the nonlinear term is on the order of $N\chi\sim 2\pi\times30\,\text{Hz}$, we have to work with Rabi frequencies of $\Omega\sim 2\pi\times20\,\text{Hz}$ to obtain $\Lambda \approx 1.5$, which makes the system sensitive to detuning on the level of less than $1\,\text{Hz}$. Slow magnetic field changes due to small temperature drifts of the magnetic field sensor translate into detuning of $\approx 10\,\text{Hz}/\text{mG}$. To obtain consistent conditions, we perform automated Ramsey experiments on the two-photon transition after 30-40 experimental repetitions and adjust the setpoint of the magnetic field stabilization accordingly. Compared to magnetic field measurements on a linear Zeeman sensitive transition, this has the advantage that small drifts of the AC Zeeman shift are also compensated. The additional information from these periodic reference measurements is used to filter out repetitions where subsequent Ramsey experiments differ by more than $1.5\,\text{Hz}$.
Measurements of the Rabi frequency are performed on a daily basis and small drifts $\le 0.5\%$ are corrected by slight adjustments of the radio frequency power.

Because of the dependence of the AC Zeeman shift on the power of the driving fields, special care was taken to ensure the resonance condition separately for all rotation pulses on the level of $\pm 1\%$ of the respective Rabi frequency. For this, we measure small amplitude Josephson oscillations with mean relative phase $\phi$ of $0$ (plasma oscillations) and $\pi$ ($\pi$ oscillations) and adjust the detuning such that they both have the same amplitude and offset in the imbalance $z$. 
The frequency difference of the same measurements is used to estimate the initial nonlinearity~\cite{ZiboldPRL2010}.

Systematic overestimation of the Fisher information from the fits to the Hellinger distance could be caused by an underestimation of the 
effective Rabi frequency during the $\theta$-rotation pulses.
To minimize effects of switching, we attenuate the radio frequency and work with an offset rotation angle of 6 degrees. Thus the smallest angle (3.5 degrees, corresponding to $\theta=-2.5$ degrees) still translates into a pulse duration of $\sim 60\,\text{\textmu s}$, which corresponds to $\sim 360$ cycles of the radio frequency. The rounding of the pulse length to full $\text{\textmu s}$ is included in the data analysis. From routinely performed measurements of the Rabi frequency, we can constrain the systematic error to $<1\%$ which translates into an uncertainty of $\Delta\theta<0.025$ degrees and is negligible compared to the error bars of the squared Hellinger distance.
\vspace{-0.3cm}
\subsection*{Experimental probability distributions}
\vspace{-0.2cm}
To obtain the experimental probability distributions for the Hellinger distance analysis, we collect the measurements of $z$ in bins of width $\Delta z = 4/N$ after postselection for the total atom number $N$. This width is slightly smaller than the experimental resolution due to photon shot-noise of $\sim\pm 4$ atoms in the two population numbers $N_a$ and $N_b$, which translates into an uncertainty of $(\Delta z)_{\text{PSN}} \approx 6/N$. By varying the bin width, we observe the expected saturation of the Hellinger distances for bin widths between $2/N$ and $\sim 5/N$. The value $4/N$ is chosen to minimize artificial broadening of essential features of the distributions while still providing sufficient statistics in each bin. To obtain the experimental probabilities, each count is divided by the total number of counts in the distribution. For the Hellinger distance measurements, we typically collect $\sim 2000$ experimental realizations at $\theta = 0$ and $\sim 500$ at $\theta \neq 0$.
\vspace{-0.3cm}
\subsection*{Reconstruction of the density matrix}
\vspace{-0.2cm}
For quantum state reconstruction, we collect the measurements of $z$ for each tomography angle $\alpha$ in bins of width $\Delta z=2/N$ according to the granularity of the symmetric Hilbert space of $J=N/2$ with discrete eigenvalues $\{m\}=\{-N/2, \dots ,N/2\}$ of $\hat{J}_z$. We then use the iterative algorithm described in~\cite{Lvovsky2004} to obtain a maximum likelihood estimate of the density matrix $\rho$. For visualization, the Husimi projection $\propto\langle\vartheta,\phi\vert\rho\vert\vartheta,\phi\rangle$ on the coherent spin
states~\cite{ArecchiPRA1972}
\begin{equation}
\label{CSS}
\vert\vartheta,\phi\rangle = \sum_{m=-J}^{J} \binom{2J}{m+J}^{1/2} \frac{\tau^{m+J}}{(1+|\tau|^2)^J}\; \vert J,m\rangle
\end{equation}
with $\tau=\exp(-i\phi)\tan(\vartheta/2)$ is calculated on a grid of $255\times 255$ points for the azimuthal angle $\phi$ and the polar angle $\vartheta$, respectively. For the density plots in \hyperlink{Fig2}{Fig.~2A} of the main text, the obtained values are divided by their maximum.
\vspace{-0.3cm}
\subsection*{Squeezing analysis}
\vspace{-0.2cm}
The separable coherent spin state (Eq.~\ref{CSS}) has the binomial variance $(\Delta z^2)_{\text{CSS}} = (4/N^2) \Delta J_z^2 = 4p(1-p)/N$ in $z$, where $p=(\langle z \rangle+1)/2$.
This is a consequence of the spin uncertainty
of $N$ atoms independently prepared in the same superposition state. We refer to the value $\xi_N^2=\Delta z^2/(\Delta z^2)_{\text{CSS}}$ as number squeezing factor
and to the value
\begin{equation}
\xi^2 = \frac{\xi_N^2}{\mathcal{V}^2} = \frac{N}{4p(1-p)\mathcal{V}^2}\Delta z^2
\end{equation}
as spin squeezing factor. $\mathcal{V} \le 1$ is the visibility of a perfect Ramsey experiment, 
which is limited by the elongation of the quantum state leading to a reduction of the 
mean spin length $\langle\vec{J}\rangle$~\cite{WinelandPRA1994}.
We estimate $\mathcal{V}$ from auxiliary interleaved measurements with the tomography angle $\alpha$ that results in 
the biggest variance of $z$ (longest axis of the state). From these measurements, we deduce the normalized mean spin length 
$\langle \cos (\pi/2-\vartheta)\rangle = \langle\sqrt{1-z^2}\rangle = \mathcal{V}$.
For the results shown in \hyperlink{Fig2}{Fig.~2C} of the main text, we average the values for $\xi^2$ at every tomography angle $\alpha$ over all 
settings of $\theta$. 
We note that reported results have not been corrected for detection or technical noise contributions. Values stated in dB are calculated as $\xi_{(N)}^2[\text{dB}]=10\log_{10}(\xi_{(N)}^2)$.
\vspace{-0.3cm}
\subsection*{Fisher Information and Cram\'er-Rao bound}
\vspace{-0.2cm}
Let $\{P_z(\theta)\} = \{P(z|\theta)\}$ be the conditional probability distribution of the random variable $Z$, which continuously depends on the parameter $\theta$, 
and $\Theta(Z)$ an unbiased estimator of $\theta$, i.e. 
$\langle\Theta\rangle = \theta$, where
$\langle \cdot \rangle$ indicates the expectation value.
We start from the equalities 
\begin{eqnarray}
&& \frac{\partial \langle\Theta\rangle }{\partial \theta} = \frac{\partial }{\partial \theta} \sum_z \Theta P_z(\theta) = 1, \\
&& \frac{\partial }{\partial \theta} \sum_z P_z(\theta) = 0.
\end{eqnarray}
With the assumption that the summing range does not depend on $\theta$, we get
\begin{equation}
\sum_z (\Theta-\theta) \frac{\partial}{\partial \theta} P_z(\theta) = \left\langle (\Theta-\theta) \frac{\partial}{\partial \theta} \log P_z(\theta) \right\rangle = 1.
\end{equation}
Taking the square of both sides, the Cauchy-Schwarz inequality 
$\langle f g \rangle^2 \leq \langle f^2 \rangle \langle g^2 \rangle$, 
where 
$f = (\Theta-\theta)$ and $g=\partial_\theta \log P_z(\theta)$,
yields
\begin{equation}
\langle(\Theta-\theta)^2\rangle \left\langle\left(\frac{\partial}{\partial \theta} \log P_z(\theta)\right)^2\right\rangle \geq 1.
\end{equation}
We thus obtain the Cram\'er-Rao bound $\Delta \Theta^2 \ge 1/F$, where $\langle (\Theta-\theta)^2 \rangle$ 
is the variance $\Delta \Theta^2$ of the estimator $\Theta$ and
\begin{equation} 
\label{FI}
F = \sum_z P_z(\theta) \left(\frac{\partial}{\partial \theta} \log P_z(\theta)\right)^2
\end{equation}
is the Fisher information.
The extension to the case of $m$ independent measurements leads to 
$\Delta \Theta^2 \ge 1/m F$~\cite{HelstromBOOK}, 
which is a natural extension obeying the extra scaling with $1/m$ for multiple measurements.
\vspace{-0.3cm}
\subsection*{Extraction of the Fisher information from the Hellinger distance}
\vspace{-0.2cm}
We first illustrate the basic theory by considering the ideal situation where 
the probabilities are known. Later we will consider the experimentally relevant case
where relative frequencies (experimental probabilities) are acquired.\\
\\*
{\it Ideal case: probabilities.}
The squared Hellinger distance between the probability distributions $p_0 \equiv \{ P_z(0) \}$ at $\theta_0=0$ and 
$p_\theta \equiv \{ P_z(\theta) \}$ at finite $\theta$ is
\begin{equation}
\label{HD}
\dHs(p_0,p_\theta) =  1 - \sum_{z} \sqrt{P_z(0) \, P_z(\theta)},
\end{equation}
where the sum, generally referred to as the Bhattacharyya 
coefficient (statistical fidelity or overlap),
extends to all values of $z$.
For small $\theta$, i.e. closely spaced probability distributions,
the Taylor expansion of the squared Hellinger distance yields
\begin{equation} 
\label{HDexp}
\dHs(p_0,p_\theta) = \frac{F}{8} \theta^2 + \frac{F'}{16} \theta^3 + \mathcal{O}(\theta^4),
\end{equation}
where $F$
is the value of the Fisher information (Eq.~\ref{FI})
at $\theta=0$ and $F'=\ud F(\theta)/\ud \theta\vert_{\theta=0}$ its derivative.
The zeroth order of the Taylor expansion vanishes 
because $\sum_z P_z(\theta) = 1$ and the first order vanishes because
$\frac{ d}{d\theta}\sum_{z} P_z(\theta)=0 $.
Thus, the Fisher information can be extracted from a polynomial 
fit to $\dHs(p_0,p_\theta)$ by extracting the coefficient of the quadratic term.
Note, that $F \geq 0$, while $F'$ and
the higher order terms of the expansion can also be negative.
By shifting the reference frame to $\theta_0 \neq 0$, this procedure can easily be generalized 
to obtain the Fisher information at arbitrary values of $\theta$. \\
\\*
{\it Experimental case: frequencies.}
The method illustrated above to extract the Fisher information is valid for any state, 
phase shift operation and measurement.
However, it requires the knowledge of the exact probability distributions.
The corresponding experimentally obtainable value is the
relative frequency (experimental probability) distribution $\{\Frq_z (\theta)\}$ for given $\theta$, 
which approaches $\{ P_z(\theta) \}$ for infinitely many independent measurements. 
We thus consider here the extraction of the Fisher information from a natural extension of the 
Hellinger distance (Eq.~\ref{HD}),  
\begin{equation} 
\label{Hellfreq}
\dHs(f_0,f_\theta) \equiv 1 - \sum_{z} \sqrt{\Frq_z(0) \, \Frq_z(\theta)},
\end{equation}
where $f_0 \equiv \{ \Frq_z(0) \}$ and $f_\theta \equiv \{ \Frq_z(\theta) \}$
are the outcome frequencies obtained from a sample of $M$ experimental realizations.
Due to statistical fluctuations, $\dHs(f_0,f_\theta)$ varies when 
repeating the measurement.
We write
\begin{equation}
\Frq_z(\theta) = P_z(\theta) + \delta \Frq_z(\theta), 
\end{equation}
where $\{P_z(\theta)\}$ is the true underlying probability distribution (from which $\{\Frq_z(\theta)\}$ is sampled) 
and $\delta \Frq_z(\theta)$ are the multinomial sampling errors characterized by
\begin{eqnarray}
\nonumber
\langle \delta \Frq_z(\theta)\rangle &=&0\\
\label{df_var}
\langle \delta \Frq_z(\theta)^2\rangle &=& \frac{P_z(\theta)(1-P_z(\theta))}{M}\\
\nonumber
\mathrm{Cov}[\delta \Frq_z(\theta),\delta \Frq_{z'}(\theta)] &=& -\frac{P_{z}(\theta) P_{z'}(\theta)}{M}\;\text{for}\;z \neq z'
\end{eqnarray}
Because of the normalization, frequency fluctuations sum to zero: $\sum_z \delta \Frq_z(\theta) = 0$.
We now expand Eq.~\ref{Hellfreq} in the Taylor series around $\theta_0=0$ and $\delta \Frq_z(\theta) \ll P_z(\theta)$.
Since $\langle \delta \Frq_z(\theta) \rangle = 0$, this is justified 
if $\sqrt{\langle \delta \Frq_z(\theta)^2 \rangle} \ll P_z(\theta)$ which, 
according to Eq.~\ref{df_var}, corresponds to the condition
$P_z(\theta) \gg 1/(M+1)$. This is satisfied for $M$ sufficiently large.
Taking the sample average, we find
\begin{equation} 
\label{averged_dh2}
\langle \dHs(f_0,f_\theta) \rangle = c_0 + 
\left( \frac{F}{8} + c_2 \right) \theta^2 + \mathcal{O}(\theta^3,\delta \Frq_z^3),
\end{equation}
with
\begin{eqnarray}
c_0 &=& \frac{n-1}{4M}\\
\nonumber
c_2 &=& \frac{1}{32M}\left[F + \sum_z \left(\partial_\theta \log P_z(\theta)\right)^2\right],
\end{eqnarray}
where $n$ is the number of discrete values of z for which $P_z(\theta)\ne 0$.
Since $F$ is the expectation value of $(\partial_\theta \log P_z(\theta))^2$ at $\theta=0$, 
we can estimate  $c_2 \approx {F (1+n)/(32M)}$.
We emphasize that  the first order term in the $\theta$-expansion of Eq.~\ref{Hellfreq} vanishes.
Notably, this happens also in the presence of frequency fluctuations.
The estimation of the Hellinger distance with experimental probability distributions is 
asymptotically unbiased. The bias decreases with $1/M$ and is due to 
the finite fluctuations on the estimated probabilities and their strict positiveness.
It can be reduced by employing a resampling procedure (see below).
Regarding the statistical error of the Hellinger distance, we find
\begin{equation} 
\label{averged_dh4}
\left( \Delta \dHs(f_0,f_\theta) \right)^2 = \frac{F}{8M} \theta^2 + \mathcal{O}(\theta^3,\delta \Frq_z^3).
\end{equation}
\vspace{-0.3cm}
\subsection*{Resampling procedure}
\vspace{-0.2cm}
Both $c_0$ and $c_2$ in Eq.~\ref{averged_dh2} scale as $1/M$, which is a prerequisite of the Jackknife procedure for bias reduction~\cite{MillerBIOMETRICA1974}.
We divide the $M$ experimental realizations in $g$ blocks of $h$ samples ($M=hg$).
For the calculation of the Jackknife estimates we use $(\dHs)_{i}$, 
that is the squared Hellinger distance 
evaluated with the $i$th group of experimental results of block size $h$ removed $(i=1,\dots,g)$.
The Jackknife estimator
\begin{equation}
\left\langle\dHs\right\rangle_\text{J} = g \dHs-\frac{g-1}{g}\sum_{i=1}^g (\dHs)_{i}
\end{equation}
has the property to eliminate the $1/M$-term from the estimate $\left\langle \dHs \right\rangle$. 
Furthermore, the variance of $\{(\dHs)_{i}\}$ is used to estimate the statistical uncertainty of $\dHs$.
The results of this block-Jackknife are averaged up to blocksize $h=20$ for mean and variance. 
We verified the validity of this approach with Monte-Carlo simulations including realistic probability distributions and typical experimental sample sizes.
\vspace{-0.3cm}
\subsection*{Bayesian estimation}
\vspace{-0.2cm}
Let $\theta_0$ be the true value of the phase shift and $\{ z_i \}_m =\{z_1,\dots,z_m\}$
a sequence of $m$ independent measurement results, obtained with probability  
$P_{\{z_i\}_m}(\theta_0) = \prod_{i=1}^m P_{z_i}(\theta_0)$.
The Bayes theorem 
\begin{equation}
P_\theta(\{z_i\}_m) = \frac{P_{\{z_i\}_m}(\theta)P(\theta)}{P(\{z_i\}_m)} 
\end{equation}
assigns a probability distribution $P_\theta(\{z_i\}_m)$ to the variable $\theta$, conditioned by the 
measurement sequence $\{ z_i \}_m$.
Here $P(\theta)$ represents the prior knowledge about the phase shift $\theta$ which is assumed to be 
phase-independent in our analysis, and $P(\{z_i\}_m)$ provides the normalization of $P_\theta(\{z_i\}_m)$.
The shape of $P_\theta(\{z_i\}_m)$ depends on the specific probabilities $P_{z_i}(\theta)$ and on the observed results. 
Nevertheless, in the limit $m \to \infty$, $P_\theta(\{z_i\}_m)$ becomes normally distributed, centered around the true value of the 
phase shift and with variance  $\sigma^2 = 1/m F$.
This result is generally referred to as the Laplace-Bernstein-von Mises theorem~\cite{LeCamBOOK}.
To demonstrate it, we rewrite $P_\theta ( \{z_i\}_m )$ in terms of frequencies $\Frq_z(\theta_0)$ to observe the result $z$:
$P_\theta ( \{z_i \}_m ) \propto \prod_z [P_\theta (z)]^{m \Frq_z(\theta_0)}$, where the proportionality sign 
indicates equality up to a normalization constant. 
In the limit $m\to \infty$, frequencies tend to probabilities ($\Frq_z(\theta_0) \to P_{z}(\theta_0)$) and thus 
\begin{equation} \label{BayPz}
\frac{\log P_\theta ( \{z_i\}_m )}{m} \to \sum_{z} P_z(\theta_0) \log P_{\theta}(z) + \text{const.},
\end{equation}
where the constant term contains the normalization of $P_\theta ( \{z_i\}_m )$.
We now expand this equation in the Taylor series around the maximum of $P_\theta ( \{z_i\}_m )$.
First, the condition $\partial_\theta \log P_\theta ( \{z_i\}_m )=0$ is satisfied if $\theta=\theta_0$.
We will assume that $\theta_0$ is the only maximum of Eq.~\ref{BayPz}.
Up to the leading order in $m$ and neglecting constant terms, which are 
absorbed in the normalization of $P_\theta ( \{z_i\}_m )$, 
we thus have 
\begin{equation}
\log P_\theta ( \{z_i\}_m ) \approx - \frac{mF(\theta_0)}{2}(\theta - \theta_0)^2.
\end{equation}
The fact that $F > 0$ guarantees that higher order terms of the expansion are negligible for $m\to \infty$.

For the Bayesian analysis presented in the main text we take advantage of the fact that the experimental
probability distribution used as $P_z(0)$ (reference) for the calculation of the Hellinger distance was collected with 
approximately four times the experimental repetitions compared to $P_z(\theta \neq 0)$. 
We randomly draw 1000 repetitions $z_i$ of this setting and use the remaining ones to generate the distribution $P_z(0)$. 
We then determine the region $a\le z\le b$, where $P_z(\theta_j)>0$ 
for all $\theta_j$. The 1000 repetitions are used as independent realizations for Bayesian estimation of $\theta$ 
and grouped in $m$-sequences $\{z_i\}_m$.
For each $m$-sequence we calculate the likelihood $\mathcal{L}(\theta_j) = \prod_{i=1}^m P_{z_i}(\theta_j)$ (equivalent to 
$P_{\theta_j}(\{z_i\}_m)$ without normalization and $P(\theta)=1$) for the (discrete) values of $\theta_j$ for which experimental 
probabilities are available. 
Realizations with $P_{z_i}(\theta_j)=0$ for some $\theta_j$, i.e. outside the range $[a,b]$ are discarded.
This is an essential step to obtain a finite value of the likelihood for each sequence. 
Note, that we do {\it not} reduce the value of $m$ in this step. 
Discarding too many realizations would thus show up as reduced sensitivity in the further analysis.

We use no prior knowledge on the phase shift, i.e. $P(\theta)=1$.
As discussed above, we expect $P_{\theta_j}(\{z_i\}_m)$ and thus the likelihood $\mathcal{L}(\theta_j)$ to be 
normally distributed around the true value of the phase shift $\theta=0$.
We thus fit the results for $\log \mathcal{L}(\theta_j)$ quadratically and extract $\sigma$ from the curvature of the fit.
As discussed in the main text, for sufficiently large $m$ (see \hyperlink{Fig4}{Fig.~4}), we get $\sigma^2 \to 1/m F$,
where $F$ agrees with the value of the Fisher information obtained with the Hellinger distance method.
\end{document}